# Biaxial strain engineering on the superconducting properties of MgB$_2$ thin film

Zhao Liu[a,b] and Biao Wang[a,b,c]*

a. School of Materials Science and Engineering, Dongguan University of Technology, Dongguan 523808, China

b. Research Institute of Interdisciplinary Science, Dongguan University of Technology, Dongguan 523808, China

c. Sino-French Institute of Nuclear Engineering and Technology, Sun Yat-Sen University, Zhuhai 519082, China

Corresponding author: B. Wang, wangbiao@mail.sysu.edu.cn

**Abstract**

The effect of biaxial strain on the superconducting properties of MgB$_2$ thin films was studied by first-principles calculations. The stability analyses by phonon dispersions show that biaxial strain as much as 7% can be applied onto MgB$_2$ without inducing any imaginary frequency. The superconducting property calculations based on the frame of Migdal-Eliashberg theory successfully reproduce the two-gap superconductivity of MgB$_2$. The results show that the tensile biaxial strain can increase the critical temperature of MgB$_2$ while the compressive biaxial strain would decrease the critical temperature. The detailed microscopic mechanism of the biaxial strain effect on the superconducting properties was studied by calculations of electronic structures and phonon dispersions. The increased $T_c$ is a combining result of the increased electron density at the Fermi level and the in-plane boron phonon softening. By means of high-throughput screening of proper substrates, it is found that most of the substrates would result in tensile strain in MgB$_2$ film, which is in agreement with many experimental works. The results in this work provide detailed understanding of the biaxial strain engineering mechanism and demonstrate that biaxial strain engineering can be an effective way of tuning the superconducting properties of MgB$_2$ and other similar materials.

**Introduction**

MgB$_2$ has been attracting consistent attention ever since the discovery of its superconductivity due to its high transition temperature $T_c$ [1, 2]. MgB$_2$ is a conventional BCS superconductor that its superconductivity is dominated by electron-phonon coupling [3, 4]. Despite the high $T_c$, extensive research has been conducted to further increase its $T_c$ [5-7]. Various methods have been tried, among which the most common techniques are mechanical loads and substitutional doping[8-10]. However, substitutional doping is not easy to achieve experimentally and can sometimes even suppress the $T_c$ [11]. In contrast, applying mechanical strain is more achievable. For example, by selecting the proper substrates for film growing, compressive or tensile strain can be introduced due to the lattice parameter mismatch between the substrate materials and MgB$_2$ [12, 13]. In addition, when synthesized as superconducting device, the influence of residual strain on the film properties needs to be considered[14, 15]. Therefore, understanding how the superconductivity of MgB$_2$ is affected by strain is important for both fundamental research and actual application.

Many researchers have successfully synthesized MgB$_2$ thin film experimentally by techniques such as hybrid physical-chemical vapor deposition (HPCVD), magnetron sputtering and molecular beam epitaxy (MBE). Among them HPCVD can prepare MgB$_2$ thin film with good quality which is critical for the investigation of the strain effect. For example, Seong *et al.* used hybrid physical-chemical vapor deposition method to grow single-crystal like MgB$_2$ thin film [16]. In spite of different preparation techniques, many have reported higher $T_c$ of the MgB$_2$ film than the bulk counterpart [17, 18]. Some efforts have been devoted to unveiling the underlying mechanism of the increased $T_c$. Pogrebnyakov *et al.* reported the systematic increase of the $T_c$ of MgB$_2$ film by HPCVD and proposed the possible mechanism is due to the $E_g$ mode softening [5]. Bekaert *et al.* found the unique three-gap superconductivity in monolayer MgB$_2$ by means of first-principles calculations [19]. Zheng *et al.* proposed a general rule that governs the enhancement (or suppression) of $T_c$ in strained MgB$_2$ [20]. These works are helpful for understanding the mechanism but how the electronic structure and phonon dispersion change under the effect of biaxial strain is lacking.

In this paper, combining the first-principles calculations and electron-phonon coupling theory based on Eliashberg theory, we investigated the biaxial strain influence on the superconductivity of MgB$_2$ thin film. How the electronic structure and phonon vibration change under biaxial strain is clearly observed. By means of high through-put screening, some new substrates for growing MgB$_2$ thin films are proposed. The strain engineering mechanism on the superconductivity of MgB$_2$ was investigated on the microscopic level, which provides reference for understanding the $T_c$ enhancement and also provides guidance for tuning the $T_c$ of other similar superconductors.

**Computational Methods**

First-principles calculations based on density functional theory (DFT) were used to investigate the effect of biaxial strain on the superconducting properties of MgB$_2$. VASP package was used for the geometry optimization for simulating the biaxial strain on MgB$_2$ [21]. PBE functional was used as implemented in VASP [22]. Monkhorst-Pack k-point grids of 9 × 9 × 9 were used for integration within the Brillouin zone. Wave functions are represented in a plane-wave basis truncated at 400 eV. The geometry relaxations were performed with a conjugate-gradient algorithm. The electronic self-consistent filed (SCF) convergence threshold was 1 × 10$^{-6}$ eV while the geometry relaxations were stopped when the forces on the atoms were below 0.01 eV·Å$^{-1}$. The biaxial strain was realized by fixing the in-plane lattice a/b while c was allowed to be relaxed [23]. The biaxial strain is defined as η = (a-a$_0$)/a$_0$, where a$_0$ is the in-plane lattice parameter of fully relaxed MgB$_2$. Periodic boundary conditions were applied along all the three directions so that possible surface effects were completely neglected.

Once the structures with different biaxial strains were obtained, their corresponding superconducting properties were calculated using the EPW package [24] combined with Quantum ESPRESSO (QE) [25]. For generating the inputs for the electron-phonon coupling calculations, electronic structure and phonon calculations were first conducted using QE. The energy cutoff of 60 Ry was used for the electronic SCF in QE with a 12 × 12 × 12 k-point grid. The exchange-correlation functional was also PBE. The plane wave basis sets for Mg and B were both from the solid-state pseudopotentials precision (SSSP) library [26]. A tolerance of 1 × 10$^{-10}$ eV was used for the energy convergence. The phonon frequencies and eigenvectors were obtained by Density Functional Perturbation Theory (DFPT) with a q-points grid of 8× 8 × 8. For calculating the $T_c$, since we focus more on the $T_c$ variation, instead of its absolute value, we used the convenient Allen-Dynes formula as implemented in EPW to calculate $T_c$ directly [27], rather than determining $T_c$ by the temperature where the superconducting gap vanishes. The formula for $T_c$ is given as

$$T_c = \frac{\omega_{log}}{1.2} \exp\left(-\frac{1.04(1+\lambda)}{\lambda - \mu^*(1+0.62\lambda)}\right) \quad (1)$$

where μ* is the pseudo Coulomb potential, which is 0.16; $\omega_{log}$ is the logarithmic averaged phonon frequency; $\lambda$ is the electron-phonon coupling strength. $\omega_{log}$ and $\lambda$ are defined as follows:

$$\omega_{log} = \exp\left(\frac{2}{\lambda}\int_0^\infty d\omega \frac{\alpha^2 F(\omega)}{\omega} \log\omega\right) \quad (2)$$

$$\lambda = 2 \int d\omega \frac{\alpha^2 F(\omega)}{\omega} \quad (3)$$

The $\alpha^2 F(\omega)$ in the equations is the Eliashberg function as defined

$$\alpha^2 F(\omega) = \frac{N(\varepsilon_F) \sum_{kvq} \left|M_{k,k+q}^{vq}\right|^2 \delta(\omega-\omega_{vq})\delta(\varepsilon_k-\varepsilon_F)\delta(\varepsilon_{k+q}-\varepsilon_F)}{\sum_{kq} \delta(\varepsilon_k-\varepsilon_F)\delta(\varepsilon_{k+q}-\varepsilon_F)} \quad (4)$$

where $N(\varepsilon_F)$ is the electronic density of states at the Fermi level $\varepsilon_F$; $\omega_{vq}$ and $\varepsilon_k$ are the phonon and electron energies, respectively, $M_{k,k+q}^{vq}$ are the electron-phonon matrix elements. The electron-phonon matrix elements were interpolated on the 12 × 12 × 12 k-point grid and 6× 6× 6 q-point grid. The visualization of the 3D Fermi surface and the superconducting gap on it was done through the Python scripts provided in the EPW package. The bonding strength between the B-B atoms was analyzed by the package of LOBSTER (Local Orbital Basis Suite Towards Electronic-Structure Reconstruction) [28].

**Results and Discussion**

Figure 1a shows the crystal structure of $MgB_2$, which possesses the hexagonal AlB2-type structure. As can be seen, the boron atoms form the 2D honeycomb graphene-like layer which is sandwiched by the two magnesium layers. The Mg-Mg bond length on the horizontal plane is equal to the lattice constant a/b and each Mg atom is at the center of a hexagonal prism of B atoms. Due to the higher electron negativity of B than Mg, the negatively charged B layer resembles the negatively charged Cu-O layer in the high-$T_c$ cuprates, which is characteristics of high $T_c$ superconductors [29]. In this work, the lattice parameters a/b and c are calculated to be 3 and 3.5 Å, respectively, which are consistent with previous theoretical results and experimental measurements [30, 31].

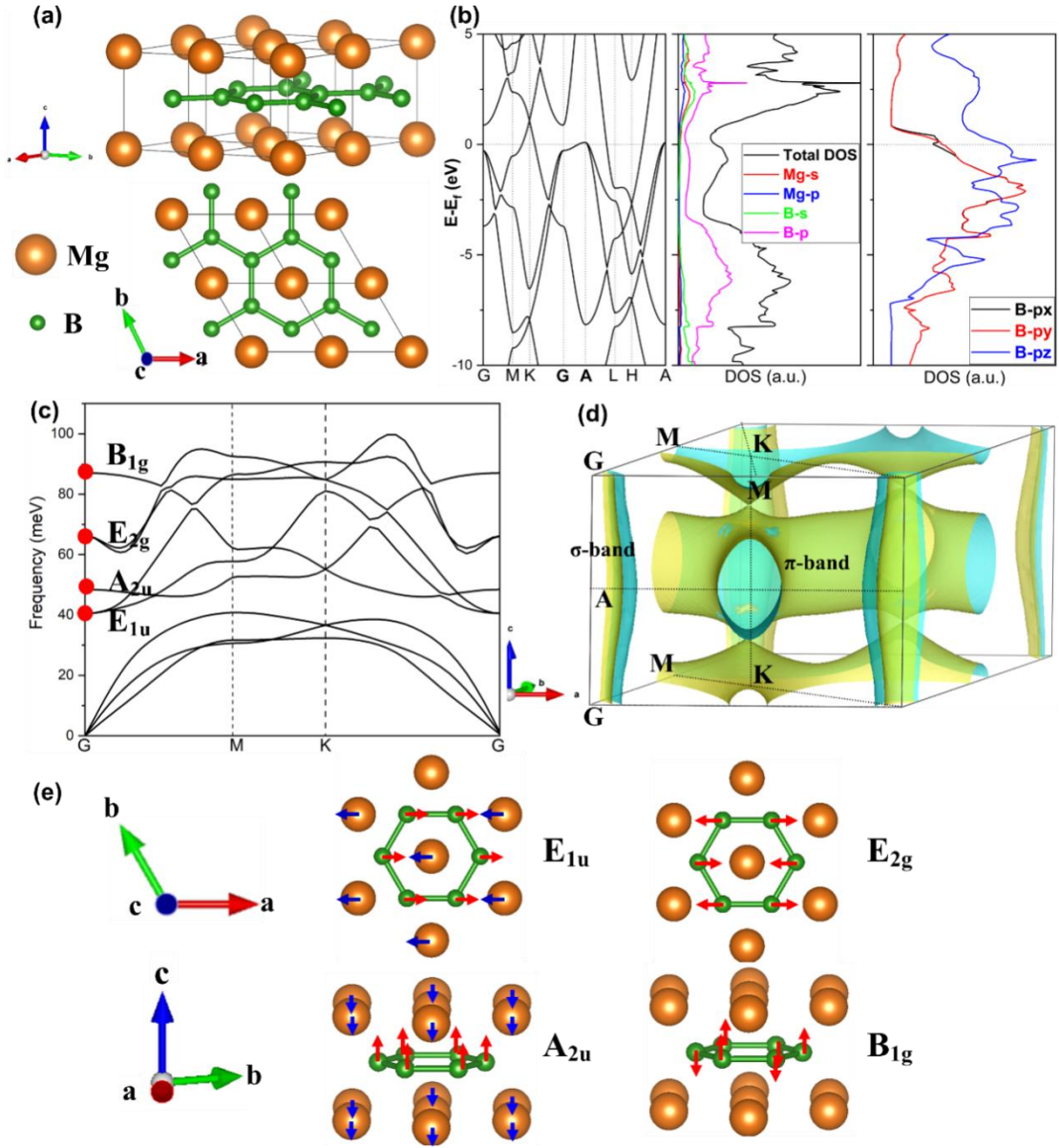

**Figure 1.** (a) The perspective view and top view of the crystal structure of $MgB_2$; (b) The electronic band structure and (P)DOS of $MgB_2$; (c) The phonon dispersion of $MgB2$; (d) The Fermi surface plot of $MgB_2$; (e) Phonon displacement patterns of $MgB_2$ at the zone center (Gamma).

For a conventional BCS superconductor like $MgB_2$, it is necessary to study its electron and phonon properties because the electrons pair through electron-phonon coupling mechanism [32]. Figure 1b shows the electronic band structure and corresponding DOS (density of states) of $MgB_2$ under strain-free state. The typical feature of the electronic band structure of $MgB_2$ is the nearly flat, double degenerate bands near the Fermi energy along the Gamma-A direction in the Brillouin zone. These bands result from the *sp*$^2$ bonding states between the B atoms, forming two $\sigma$ bonds, which give rise to nearly cylindrical, hole-like Fermi surfaces around the Gamma point [33]. Unlike graphite, which has the similar honeycomb C layer, the $\sigma$ bands in MgB$_2$ cross the Fermi surface, indicating the hole carriers dominate its transport properties, as has been demonstrated by the experimental measurement of a negative Hall coefficient [34]. The corresponding DOS suggest that MgB$_2$ is metallic for the DOS curve cross the Fermi level without any gap [35]. The partial density of states (pDOS) from the *s*, *p* orbitals of Mg and B show that the DOS contribution near the Fermi level mainly comes from B atoms, which suggest that the B layers in MgB$_2$ have more influence than Mg on its superconducting properties. Another point needs to mentioned is that the DOS near the Fermi level is relatively low compared with other BCS superconductors, which implies that tuning the phonon properties may have more impact than tuning the electronic properties on the electron-phonon coupling strength of MgB$_2$. The rightmost figure in Figure 2b is the pDOS of the three *p* orbitals of B atoms. As can be seen, the $p_x$ and $p_y$ orbitals are almost completely overlapping, while the $p_z$ orbital is separated, suggesting that two of the three original triply degenerated *p* orbitals of B atom form the doubly degenerate $\sigma$ bonds while the third *p* orbital form the $\pi$ bond. All these analyses of the electronic structures are correctly reflected on the Fermi surface plot as shown in Figure 1d. Firstly, the Fermi surface is mainly contributed by the B electrons. Secondly, the doubly degenerate $\sigma$ bonds form the two fluted cylinders near the Gamma point surrounding the Gamma-A line of the Brillouin zone [36], while the $p_z$ orbital forms the $\pi$ band.

Figure 1c shows the phonon dispersion of MgB$_2$ under zero strain. For a 3-atom unit cell, there are totally 9 vibrational modes, with 3 acoustic modes and 6 optical modes. For the 6 optical modes which can couple with electrons, they are divided into 4 groups due to the double degeneration of $E_{2g}$ and $E_{1u}$ modes. Owing to the light atomic mass of the B atoms and the strong B-B coupling, the two high-frequency modes almost have a pure boron character. The in-plane stretching $E_{2g}$ and the out-of-plane $B_{1g}$ mode (where the atoms move in opposite directions) are the boron atom modes. $E_{2g}$ is a Raman active mode and experimental studies showed that this mode is sensitive to structural changes[5]. The low-frequency modes $A_{2u}$ and double degenerate $E_{1u}$ are infrared active and are less sensitive to structural changes because they do not involve B-B in-plane bond.

Figure 2a shows the schematic representation of the $MgB_2$ film epitaxially grown on the substrates with different lattice constants. Those substrates with larger lattice constants than the in-plane lattice of $MgB_2$ would result in tensile stain in $MgB_2$ film while those substrates with smaller lattice constants than $MgB_2$ would result in compressive strain. In most cases, the substrates for growing $MgB_2$ film should be compatible in terms of their crystal structures. Up to now, epitaxial $MgB_2$ film has been successfully grown on various substrates, including $Al_2O_3$ [4, 12, 37], Mg [38], MgO [39], SiC [40], sapphire[41], $LaAlO_3$ [42], $SrTiO_3$ [43], $ZrO_2$ [42], $SiO_2$ [44]. As for the synthesis technique, the hybrid physical-chemical vapor deposition (HPCVD) and molecular beam epitaxy (MBE) are the most commonly used methods.

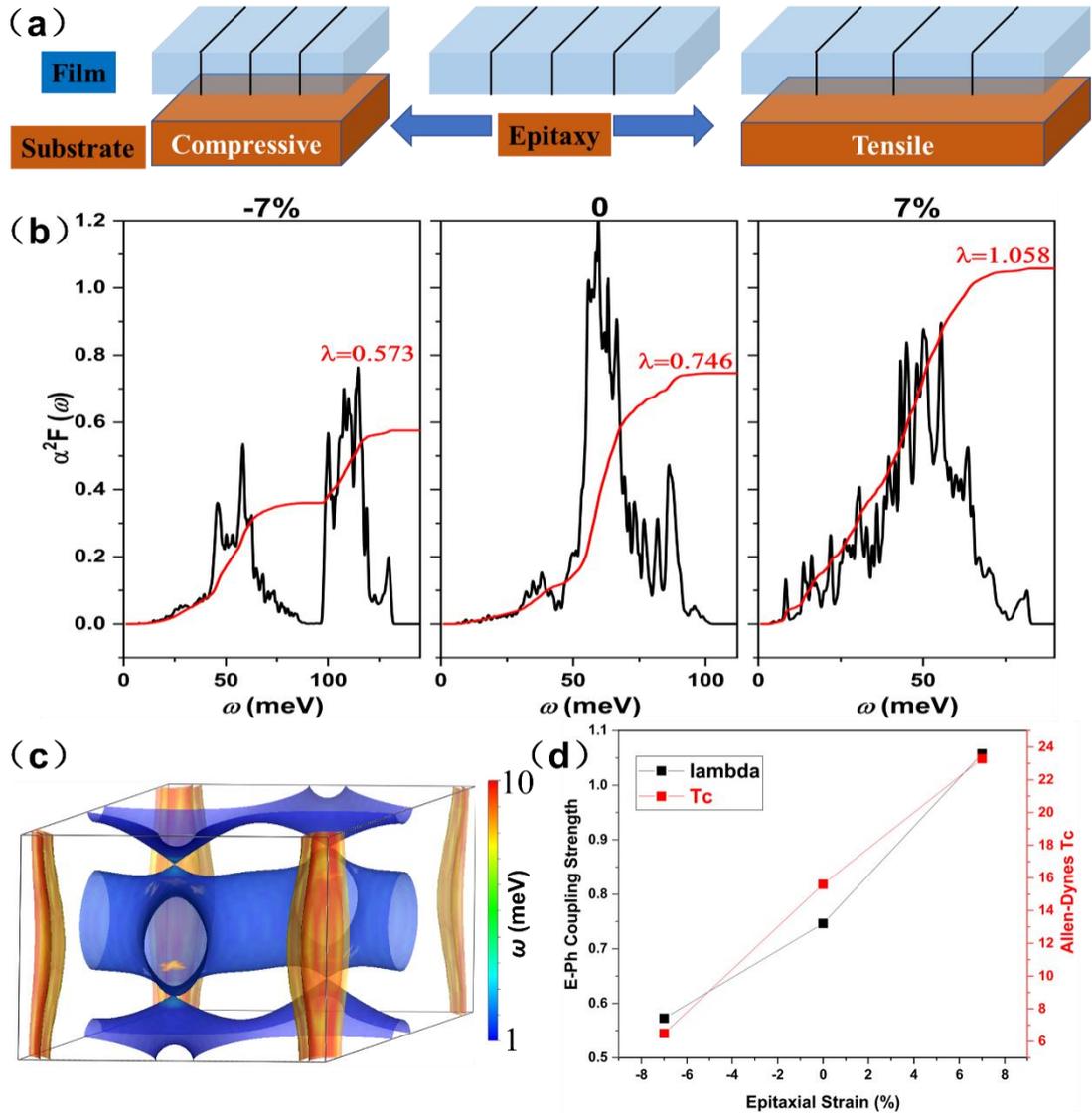

**Figure 2.** (a) Schematic diagram of the epitaxial growth of MgB$_2$ on substrates with compressive and tensile biaxial strain;(b) The calculated Eliashberg spectral function $α^2F(ω)$ of MgB$_2$ (black line) and the integrated electron-phonon coupling strength $λ$ (red line); (c) The superconducting energy gaps of MgB$_2$ (unit in meV) on the Fermi surface for T = 10 K (Colour rendering was done using VESTA software); (d) The electron-phonon coupling strength and the Allen-Dynes critical temperature as a function of the biaxial strain.

Figure 2c shows the calculated Eliashberg spectral $α^2F(ω)$ and the corresponding integrated electron-phonon coupling strength $λ$ of MgB$_2$ under different strain cases. For the case with no strain applied, the electron-phonon coupling strength is calculated to be 0.746 which is in agreement with previous studies [45]. The spectral function shows the dominant peaks ranging from 60 to 75 meV and the secondary peaks around 85 meV, which correspond to the phonon energies of $E_{2g}$ mode and $B_{1g}$ mode, respectively. As discussed above, since these two vibrational modes are mainly contributed by the B-B in-plane bond, these two vibrational modes are sensitive to biaxial strain which would leads to the in-plane structural changes. Consequently, strain engineering on the B-B in-plane bond can be an efficient way to tune the electron-phonon coupling strength of $MgB_2$, thus to tune its critical temperature.

To study the mechanism of strain engineering of $MgB_2$ on its $T_c$, we calculated the electronic and phonon structures separately. Figure 3a and 3b show the band structure and DOS of $MgB_2$ under different biaxial strains, respectively. According to the band structures, the general features of them don't change significantly under different strains. Within the studied energy range from -10 to 10 eV, the compressive strain results in wider band occupation, compared with the unstrained case, while the tensile strain results in narrower band occupation. This can be observed more intuitively in the DOS that the compressive strain leads to the DOS expansion while the tensile strain leads to the DOS compression. Therefore, $MgB_2$ has the highest electron density near its Fermi energy when under tensile biaxial strain. In contrast, $MgB_2$ has the lowest electron density near its Fermi energy when under compressive biaxial strain. In addition to the electron density variation near the Fermi energy, it is noted that the Van Hove singularity [46] formed by the B-$p$ orbital moves consistently closer to the Fermi level from the compressive strain case to the tensile strain case. These combined effects coming from the biaxial strain would lead to a strong boost of the electron-phonon coupling in the tensile strained $MgB_2$.

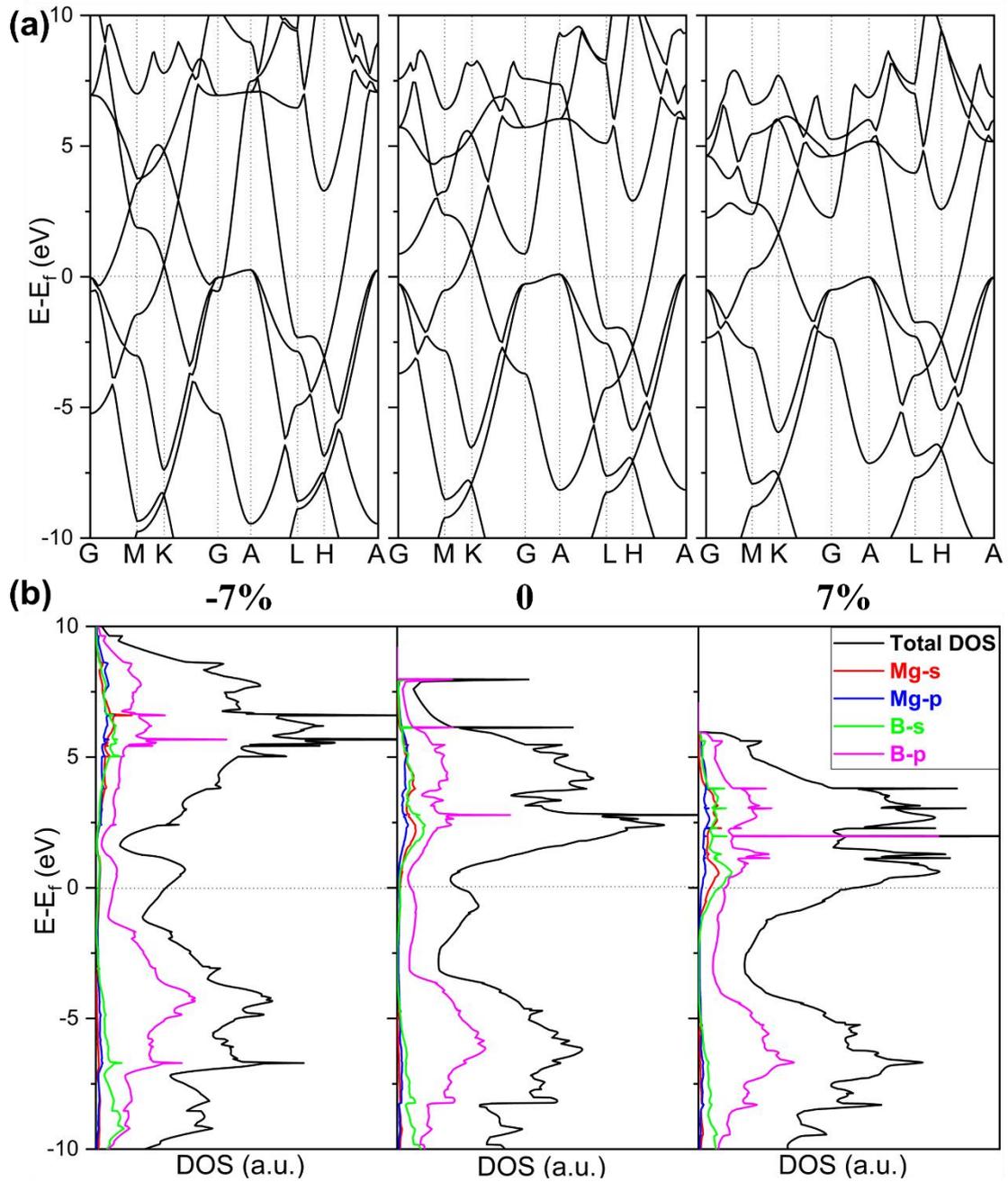

**Figure 3.** (a) The band structures of MgB$_2$ under different biaxial strains; (b) The corresponding DOS of MgB$_2$ under different biaxial strains

Singularity moving closer to Ef with tensile strain

Apart from the electronic structure effect, the effect of biaxial strain on the phonon structure is also important for understanding the strain engineering mechanism. Figure 4a shows the evolution of the lattice parameters of MgB$_2$ cell and the B-B bond length under different biaxial strains. The studied bonds are indicated on right panel of

Figure 4a. With the biaxial strain varies from the -7% compression to +7% tension, the lattice parameter $a/b$ increases monotonously from ~2.85 to ~3.25 Å, while the lattice parameter $c$ decreases monotonously from ~3.60 to 3.45 Å. This leads to the size increase on the horizontal plane and a size decrease along the perpendicular direction. Consequently, the B-B bond length also increases monotonously from the compressive side to the tensile side. Figure 4b shows the phonon dispersions of $MgB_2$ under the 7% compressive strain and the 7% tensile strain. As can be seen, there is not any imaginary frequencies, which confirms the dynamical stability of $MgB_2$. This suggests that biaxial strain of as high as 7% can be exerted on $MgB_2$ without causing structural instability. Furthermore, the frequencies of all the vibration modes consistently increases due to the compressive biaxial strain while the tensile strain results in frequency decrease. In addition, the vibration frequencies of the four optical modes confirm that the $E_{2g}$ and $B_{1g}$ modes are sensitive to the biaxial strain while the $A_{2u}$ and $E_{1u}$ are not, because their frequencies only slightly change, in contrast to the significant change of $E_{2g}$ and $B_{1g}$ modes. All these results are consistent with the above structural analyses that the B-B in-plane modes are sensitive to structural changes while the other two modes are not sensitive. The $B_{1g}$ mode is the shear vibration of the Mg and B planes, thus the change of the lattice parameters or the B-B bond length barely affect its vibration frequency. The $A_{2u}$ mode slightly increases under compression and slightly decreases under tension can be explained by the structural hardening and softening, respectively.

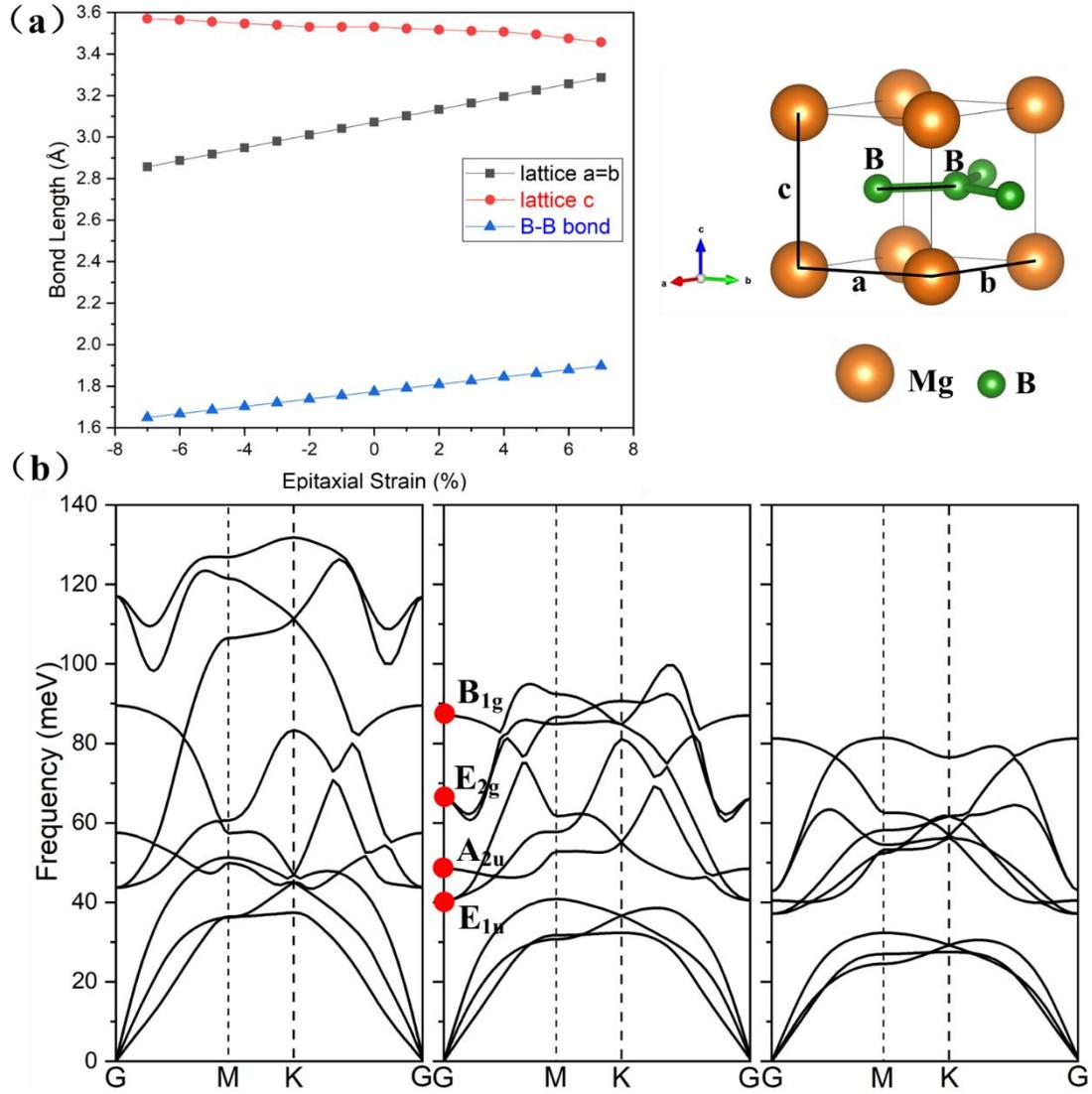

**Figure 4.** (a). Bond length evolution of $MgB_2$ under different biaxial strains, the figure on the right indicates which of the lengths are studied; (b)The phonon dispersions of $MgB_2$ under different biaxial strains.

Based on the detailed electronic structure and phonon analyses, it is now easy to explain the biaxial strain induced critical temperature change on $MgB_2$. According to the BCS theory and the subsequent refinements [27], the high $T_c$ in a phonon-mediated superconductor is governed by the large DOS at the Fermi level and the high phonon frequency. In most cases, the large electron-phonon coupling strength λ is directly related to high $T_c$. According to the McMillan-Hopfield expression [47, 48], the electron-phonon coupling strength λ can be represented in the following equation:

$$\lambda = \frac{N(E_F)<I^2>}{M<\omega^2>} \quad (5)$$

where $N(E_F)$ is the electron density of states at the Fermi level, the $<I^2>$ is the electron-phonon matrix element, $<\omega^2>$ is the averaged square of the phonon frequency, M is the mass of ion involved. According to the electronic DOS and phonon calculations presented above, when compressive biaxial strain is applied onto $MgB_2$, its DOS at the Fermi level decreases, accompanied by the increase of its optical phonon frequencies, thus leads to the decrease of the electron-phonon coupling strength. In contrast, when tensile biaxial strain is applied, the DOS at the Fermi level increases while the phonon frequencies decrease. Therefore, the electron-phonon coupling strength increases and hence the critical temperature.

Based on the Eliashberg spectral in Figure 2, the electron-phonon coupling is mainly contributed by the phonons in the range of 50 – 100 $\omega$eV, i.e., the $E_{2g}$ and $B_{1g}$ modes, which are the vibrations of the B-B bonds. To further analyze the bonding state between the B-B bonds, the Crystal Orbital Hamilton Population (COHP) analysis was conducted on the two boron atoms under different biaxial strains. COHP is a useful tool for investigating the interaction strength between two selected atoms. It projects the band-structure energy into the orbital-pair interactions. A more chemically speaking, it is a 'bond-weighted' DOS between a pair of atoms [49]. A negative COHP value indicates bonding interactions while a positive COHP value indicates antibonding interactions. By doing integral of the COHP (ICOHP), the bonding strength can be quantitively compared. Figure 5a shows the COHP of B-B atoms and their corresponding total DOS and pDOS under the three different strain states. The negative ICOHP values labeled in the figure suggest that the B-B atoms are stably bonded in all the strain states, which is in good agreement with the above dynamical stability results. The strain-free $MgB_2$ possesses the most negative ICOHP of -12.69, suggesting that it has the strongest bonding strength, followed by the compressive strained case (-10.07) and tensile strained case (-7.56). The strain-free state results in the strongest bond strength is because the geometry relaxation tends to find the energetically favorable structure which favors stronger bond strength. When the structure is fully relaxed, the two boron atoms are at their balanced positions, thus has the highest bond strength. When compressive biaxial strain is applied, the B-B distance is shortened according to Figure 4a, which is no longer at their balanced positions, and hence has weaker bond strength. However, since the bond length is decreased, leading to the hardening of the structure, thus the phonons increase. As for the tensile strained case, the B-B bond is elongated, which leads to further decrease of the bond strength compared with the compressed case. Therefore, the tensile biaxial strain results in the lowest B-B bond strength among the three cases. Figure 5b shows the electron localization function (ELF) of $MgB_2$ under different biaxial strain applied. ELF is a measure of the possibility of finding an electron in the neighborhood space and is usually used to analyze the interatomic interaction. Large ELF value indicates high localization state while small ELF value indicates non-localization state. The ELF in Figure 5b shows that the electrons are localized between the B atoms while there's almost no electron surrounding the Mg atoms, suggesting the charge transfer from Mg to B atom. This is consistent with our above analyses that the superconducting properties of $MgB_2$ is dominated by the electronic structure of boron atoms.

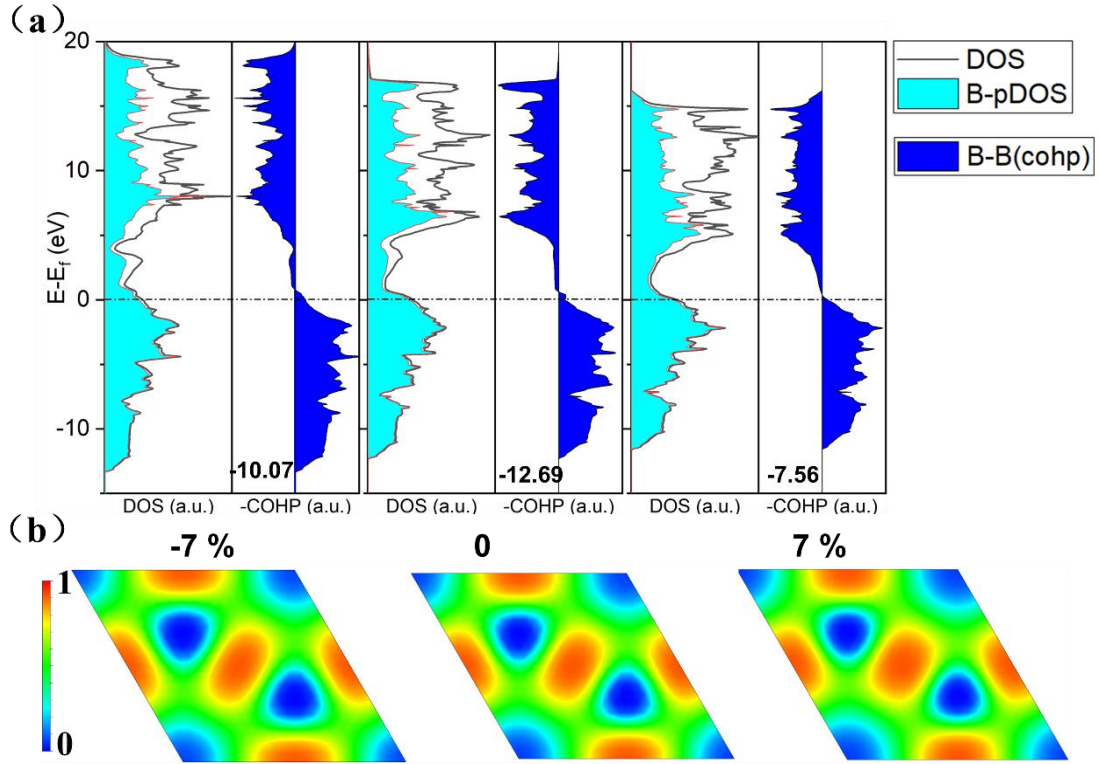

**Figure 5.** (a) Total DOS and pDOS and the corresponding crystal orbital Hamilton populations (COHP) between B-B atoms in the structure; (b) The electron localization function of MgB$_2$ under -7%, 0% and 7%, respectively.

Table 1 lists some of the commonly used substrate materials for growing the <001> orientation MgB$_2$ thin films. Some of these substrates have been reported by previous experimental work, the rest are selected from the Materials Project database, in which the selection criteria were based on the formation energy, elastic strain energy and topological information, as has been described in work of Ding *et al.* [50]. The lattice parameters for these substrates with the proper orientation are also listed, with which the resultant biaxial strain of MgB$_2$ film can be calculated accordingly. The magnitude of the resultant strain ranges from ~0.6% to ~80%, although the actual strain in practice depends on many factors, such as the synthesis technique, thickness, quality of film, *etc*. Nevertheless, all the candidate substrates listed in Table 1 would lead to tensile strain in MgB2 thin film. This explains that most of the successfully grown epitaxial MgB$_2$ films by experimental efforts have higher $T_c$ than their bulk counterparts, which is owing to the enhanced electron-phonon coupling strength by the increased electron density at the $E_f$ as well as the bond stretching mode softening. So, measuring the electron density at the $E_f$ and vibration frequencies of selected modes are helpful for unveiling the mechanism of superconductivity engineering for experimentalists.

**Table 1.** The candidate substrate materials for growing biaxial $MgB_2$ <001> thin film.

| Substrates | Orientation | Lattice (Å) | Strain (%) |
|---|---|---|---|
| SiC | <100> | 3.094 | 0.650618 |
| WS2 | <100> | 3.191 | 3.806116 |
| MoS2 | <100> | 3.191 | 3.806116 |
| Mg | <100> | 3.203 | 4.196487 |
| GaN | <100> | 3.216 | 4.619388 |
| Te2Mo | <100> | 3.559 | 15.77749 |
| CeO2 | <111> | 3.866 | 25.76448 |
| Si | <111> | 3.867 | 25.79701 |
| GaAs | <111> | 4.066 | 32.27066 |
| Ge | <111> | 4.075 | 32.56344 |
| CdS | <100> | 4.207 | 36.85751 |
| ZnSe | <111> | 4.604 | 49.77228 |
| Al2O3 | <111> | 4.801 | 56.18087 |
| LaAlO3 | <111> | 5.411 | 76.02472 |
| CsI | <111> | 5.557 | 80.77424 |

**Conclusions**

In this work, first-principles method based on density functional theory was performed to study the biaxial strain effect on the superconducting properties of $MgB_2$. The results predict that biaxial strain of as high as 7% can be exerted onto $MgB_2$ without causing dynamical instability, whereas most of the previous theoretical works only consider strain of less than 5%. Our calculations successfully reproduced two superconducting gaps on the Fermi surface. The tensile biaxial strain would result in the increase of the $T_c$, while the compressive strain would result in the decrease of the $T_c$. The underlying mechanism of the strain effect on the superconductivity was analyzed by detailed calculations of the electronic structures and phonon dispersions. The electronic analyses suggest that the DOS of $MgB_2$ near the Fermi level is mainly contributed by the boron atoms. The $\sigma$ band and $\pi$ band on the Fermi surface are formed by the double degenerate B $p_x$ and $p_y$ orbitals and the B-$p_z$ orbital, respectively. In addition, the electron-phonon coupling strength calculation based on the Eliashberg spectral demonstrate that the $E_{2g}$ and $B_{1g}$ phonon modes contribute significantly to the coupling strength. These two modes are the optical phonon branches involving only the B-B bond. Therefore, the results suggest that tuning the properties of the boron bond is the key to tune the superconducting properties of $MgB_2$. Some of the common candidate substrate materials for growing $MgB_2$ films are investigated and all of them can result in tensile strain. This explains that most of the experimentally reported $MgB_2$ films possess higher $T_c$ than its bulk counterpart, which is because the tensile biaxial strain can lead to higher electron-phonon coupling. The present work provides a detailed mechanism of the effect of biaxial strain on the superconductivity of $MgB_2$ on the microscopic level, which could be a good reference for the understanding of the strain engineering mechanism on other similar two-dimensional superconducting materials.

**Declaration of competing interest**

The authors declare no competing financial interests.

**Acknowledgments**

This work was supported by National Natural Science Foundation of China (12002402, 11832019) and and the NSFC original exploration project (12150001). The authors also appreciate the financial support from the Project of Nuclear Power Technology Innovation Center of Science Technology and Industry for National Defense (HDLCXZX-2019-ZH-31). This work was also supported by the Guangdong International Science and Technology Cooperation Program (2020A0505020005). Z L also wants to thank the financial support from the Guangdong overseas young postdoctors recruitment program.